# Studying the Prevalence of Exception Handling Anti-Patterns

Guilherme Bicalho de Pádua
Department of Computer Science and Software Engineering
Concordia University - Montreal, QC, Canada
Email: g_bicalh@encs.concordia.ca

Weiyi Shang
Department of Computer Science and Software Engineering
Concordia University - Montreal, QC, Canada
Email: shang@encs.concordia.ca

*Abstract*—**Modern programming languages, such as Java and C#, typically provide features that handle exceptions. These features separate error-handling code from regular source code and are proven to enhance the practice of software reliability, comprehension, and maintenance. Having acknowledged the advantages of exception handling features, the misuse of them can still cause catastrophic software failures, such as application crash. Prior studies suggested anti-patterns of exception handling; while little knowledge was shared about the prevalence of these anti-patterns. In this paper, we investigate the prevalence of exception-handling anti-patterns. We collected a thorough list of exception anti-patterns from 16 open-source Java and C# libraries and applications using an automated exception flow analysis tool. We found that although exception handling anti-patterns widely exist in all of our subjects, only a few anti-patterns (e.g. *Unhandled Exceptions*, *Catch Generic*, *Unreachable Handler*, *Over-catch*, and *Destructive Wrapping*) can be commonly identified. On the other hand, we find that the prevalence of anti-patterns illustrates differences between C# and Java. Our results call for further in-depth analyses on the exception handling practices across different languages.**

## I. INTRODUCTION

Exception handling features, such as throw statements and try-catch-finally blocks, are widely used in modern programming languages. These features separate error-handling code from regular code and are proven to enhance the practice of software reliability, comprehension, and maintenance [1], [2]. On the other hand, the misuse of exception handling features can cause catastrophic failures [3]. A prior study shows that two-thirds of the studied system crashes were due to exceptions [4]. Barbosa *et al.* [5] illustrate the importance of the quality of exception handling code. Similar findings were also discussed in a prior survey [6].

To improve the quality of exception handling, prior research has reported a slew of anti-patterns on exception handling. These anti-patterns describe the problematic exception handling source code that may exist in the entire life cycle of exceptions, i.e., the propagation of the exception, the flow of the exception and the handling of the exception. Although these anti-patterns are discussed in prior research [7], the prevalence of these anti-patterns is not studied in-depth.

In this paper, we investigate the prevalence of exception handling anti-patterns in 16 open-source Java and C# applications and libraries. We find that all of the studied subjects have exception handling anti-patterns detected in their source code.

Whereas only five anti-patterns (*Unhandled Exceptions*, *Catch Generic*, *Unreachable Handler*, *Over-catch*, and *Destructive Wrapping*) are prevalently observed, i.e., in median detected in over 20% of the catch blocks or throws statements in the subject systems. We observe that these anti-patterns are often associated with multiple flows of exception, leading to bigger impact and more challenging resolution of such anti-patterns. By further investigation, we find that programming languages (e.g., Java or C#) may have a relationship to the existence of anti-patterns, while we do not observe such relationship with the type of projects (e.g., application or library).

Our results imply that, despite the prior research on exception handling, there is still lacking a deep understanding of the practice of exception handling. More in-depth analyses are needed to ensure the quality and usefulness of exception handling in practice.

The rest of the paper is organized as follows: Section II presents the background of exception handling features and their anti-patterns. Section III presents our case study of exception handling anti-patterns. Finally, Section IV concludes the paper and discusses potential future research directions based on our early researching findings.

## II. BACKGROUND AND RELATED WORK

### A. Anti-patterns of exception handling

There are different actions and their respective programming mechanisms involved in exception handling: 1) defining an exception using a type declaration, 2) raising an exception using a throw statement, 3) propagating an exception in a method by not handling it or using a throws statement and 4) handling an exception using a catch block. According to the implementation of the above actions, there can be different anti-patterns. In this paper, we focus on the actions of propagation and handling of exceptions from the perspective of the explicit mechanisms (i.e. try-catch and throws). In particular, there exist three categories of related anti-patterns (see Table I): *1.Flow* anti-patterns are in the intersection of propagation (i.e. methods in the try block and its thrown exceptions) and handling actions (i.e. the catch block content) [3], [8], [9], [7]. *2.Handler* anti-patterns are only in the handling actions and are not related to the propagated exceptions [2], [3], [8], [9], [10]. *3.Throws* anti-patterns are related to propagation issues, and they are specifically related to throws statement [9], [10].

*B. Empirical studies on exception handling*

Prior research studied exception handling based on source code and issue tracking systems. Cabral and Marques [11] studied exception handling practices from 32 projects in both Java and .Net without considering the flow of exceptions or anti-patterns. Sena *et al*. [7] investigated 656 Java libraries for exceptions flow characteristics, handler actions, and handler strategies. Their study considered a smaller set of anti-patterns that were only evaluated manually in the sampled code of a sample of libraries. Sinha *et al*. [9] leveraged exception flow analyses to study the existence of 11 anti-patterns in four Java systems, without studying their prevalence. To understand the impact of anti-patterns, prior studies [2], [6], [5] classified exception-handling related bugs by mining software issue tracking. This paper is the first work to study the prevalence of exception handling anti-patterns extensively.

## III. CASE STUDY

*A. Subject projects*

Table II depicts the studied subject projects. All subject projects are open source projects obtained from GitHub. We selected subject projects (see Table II) by considering their number of stargazers and contributors.

*B. Detecting exception handling anti-patterns*

We detected all the exception handling anti-patterns presented in Table I. In particular, we leverage Eclipse JDT and .NET Compiler Platform ("Roslyn") to parse the Java and C# source code, respectively. To precisely detect these anti-patterns, we not only parse the try-catch blocks but also analyze the flow of the exceptions. Our exception flow analysis collects the possible exceptions from four different sources: documentation in the code syntax, documentation for third party and system libraries, explicit throw statements, and binding information of exceptions (not available for C#).[1]

*C. The prevalence of exception handling anti-patterns*

Our goal is to put in perspective the existence of exception handling anti-patterns. We collected source code information from a diverse set of subject projects in different programming languages. The knowledge of the prevalence of anti-patterns would help developers improve exception handling practices.

In total, we detected 17 exception handling anti-patterns from the perspective of the catch block, i.e., whether each catch block contains an anti-pattern. We also detected two exception handling anti-patterns from the perspective of the throws statements. Throws statements are used to indicate the propagation of exceptions explicitly. Since this feature is not available in C#, we only detect throws level anti-patterns in the Java projects.

**All anti-patterns are detected at least once in subject projects, while only a small amount of anti-patterns are**

[1]Source code, binaries and Tableau visualizations with raw data are available online at https://guipadua.github.io/icpc2017.

**prevalent.** As shown in Tables III and IV, all anti-patterns exist in our subject projects. In fact, the least found anti-pattern, *Incomplete Implementation*, can still be found in six projects. This finding implies that prior research indeed captures anti-patterns that correspond to the smell in practice. The existence of all anti-patterns shows the lack of awareness to the importance of quality exception handling code.

On the other hand, we find that only a small number of anti-patterns are prevalent. In particular, only five anti-patterns, i.e., *Unhandled Exceptions*, *Catch Generic*, *Unreachable Handler*, *Over-catch* and *Destructive Wrapping*, are detected in over 20% (40.8%, 31.9%, 28.0%, 24.6%, 22.3%, respectively) of the catch blocks or throws statements in median. On the other hand, all other anti-patterns are rather rare in the source code. Yuan *et al*. [3] claimed that three exception handling anti-patterns (*Over-catch and Abort*, *Catch and Do Nothing* and *Incomplete Implementation*) could cause catastrophic system failure, while we find that all these three anti-patterns are rarely detected. There are only 12 *Incomplete Implementation* anti-pattern instances detected in all the studied projects. Another surprising finding is that the most widely detected anti-pattern is *Unhandled exceptions*. This anti-pattern has been known as the common root-cause of system crashes [2], and prior research has proposed techniques to help identify all possible exceptions [12], [9]. However, our results imply that developers still overlook the importance of this anti-pattern and it may lead to potential crash at system run-time.

*D. The amount of exception flows*

The anti-patterns can be related to a single, multiple, or no exception flow at all (e.g. *Unreachable Handler*). We aim to study the number of flows affected by those anti-patterns. The larger the quantity of flows, the larger the impact of those anti-patterns.

**Multiple flows are impacted by each anti-pattern.** Table V depicts the quantity of affected flows for the flow-based anti-patterns. For *Unhandled Exceptions* and *Unreachable Handler*, 83% (C#) and 67% (Java) of the affected catch blocks have multiple impacted (uncaught) flows, with a maximum of 37 flows. For *Over-catch* and *Over-catch and Abort*, 84% (C#) and 60% (Java) of the affected catch blocks have multiple impacted (over-caught) flows, with a maximum of 43 flows.

*E. Discussion*

In this subsection, we aim to understand the existence of anti-patterns from different perspectives.

**Programming languages.** The prevalence of exception handling anti-patterns can vary between Java and C# (see Table III). Figure 1 presents examples of anti-patterns that have a large difference in prevalence between Java and C#. The box plots represent the distribution of percentages of catch blocks that contain anti-patterns in each project. For example, the median value of *Destructive Wrapping* in Java (31.6%) is almost 18 times bigger than in C# (1.8%). Another example is *Catch Generic*, in which the minimum value (45.0%, median: 74.3%) in C# is 33% higher than the maximum value (33.8%, median: 17.6%) in Java. The reason of such differences can be the nature of different exception handling strategies in C# and Java. Java forces that certain kinds of exception (i.e. *Checked* exceptions) are handled or explicitly propagated before compilation while C# does not. To support that, popular Java IDEs suggest the exceptions that should be handled. For example, if a developer adds a function call to read a file, the IDE will propose that the non-generic exception *IOException* should be handled or propagated.

TABLE I
LIST OF THE DETECTED ANTI-PATTERNS.

| Group | Anti-pattern | Short Description | Group | Anti-pattern | Short Description |
|---|---|---|---|---|---|
| Flow | Over-catch | The handler catches multiple different lower-level exceptions [8], [7]. | Handler | Incomplete Implementation | The handler only contains TODO or FIXME comments [3]. |
| | Over-catch and Abort | Besides over-catching, the handler aborts the system [3]. | | Log and Return Null | Besides being a dummy handler, the handler return null [10]. |
| | Unhandled Exceptions | The handler does not catch all possible exceptions [9]. | | Log and Throw | The handler logs some information and propagates the exception [10]. |
| | Unreachable Handler | The handler does not catch any possible exception [9]. | | Multi-Line Log | The handler divides log information into multiple log messages [10]. |
| Handler | Catch and Do Nothing | The handler is empty [3], [2], [9]. | | Nested Try | The handler and its try block is enclosed in another try block [2]. |
| | Catch and Return Null | The handler contains return null [10], [2]. | | Relying on getCause() | The handler contains a call to getCause() [10]. |
| | Catch Generic | The handler catches a generic exception type (e.g. Exception) [10], [9], [8]. | | Throw within Finally | The handler is followed by a finally block that propagates exceptions [10]. |
| | Destructive Wrapping | The handler propagates the exception as a new exception [10] | Throws | Throws Generic | The throws propagates a generic exception type (e.g. Exception) [10]. |
| | Dummy Handler | The handler only display or log some information [2]. | | Throws Kitchen Sink | The throws propagates multiple exceptions [10]. |
| | Ignoring InterruptedException | The handler catches InterruptedException and ignores it [10]. | | | |

TABLE II
OVERVIEW OF THE SELECTED SUBJECT PROJECTS.

| | Project | Release Version | Type | # Throws | # Catch | # Method (K) | KLOC |
|---|---|---|---|---|---|---|---|
| C# | Glimpse | 1.8.6 | App. | - | 57 | 1 | 31 |
| | Google API | v1.15.0 | Lib. | - | 30 | 16 | 628 |
| | OpenRA | release-20160508 | App. | - | 143 | 7 | 125 |
| | ShareX | v11.1.0 | App. | - | 341 | 7 | 177 |
| | SharpDevelop | 5.0.0 | App. | - | 1060 | 41 | 923 |
| | SignalR | 2.2.1 | Lib. | - | 105 | 2 | 38 |
| | Umbraco-CMS | release-7.5.0 | App. | - | 615 | 15 | 362 |
| Java | Apache ANT | rel/1.9.7 | App. | 1,622 | 1139 | 11 | 158 |
| | Eclipse JDT Core | I20160803-2000 | Lib. | 1,686 | 1655 | 25 | 383 |
| | Elasticsearch | v2.4.0 | App. | 1,782 | 408 | 12 | 108 |
| | Guava | v19.0 | Lib. | 509 | 317 | 10 | 79 |
| | Hadoop Common | rel/release-2.6.4 | Lib. | 4,495 | 1144 | 14 | 147 |
| | Hadoop HDFS | rel/release-2.6.4 | App. | 1,538 | 586 | 4 | 44 |
| | Hadoop MapReduce | rel/release-2.6.4 | App. | 1,221 | 367 | 6 | 57 |
| | Hadoop YARN | rel/release-2.6.4 | Lib. | 4,146 | 1529 | 29 | 257 |
| | Spring Framework | v4.3.2.RELEASE | Lib. | 5,856 | 2301 | 30 | 349 |
| | | | Total | 22,855 | 9,446 | 141 | 1,582 |

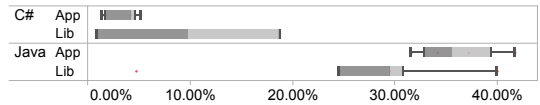

(a) *Destructive Wrapping*

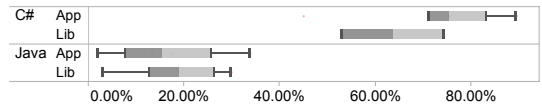

(b) *Catch Generic*

Fig. 1. Examples of differences between Java and C# and between applications and libraries. Differences between Java and C# are significant based on Wilcoxon Rank Sum test ($p$-value $<0.05$). Based on the same test, all differences between applications and libraries are not statistically significant.

**Types of projects.** Library and application projects may have different exception handling practice, where, intuitively, libraries would propagate exceptions and applications handle exceptions. We examine whether such difference impacts the prevalence of exception handling anti-patterns. Figure 1 presents examples of anti-patterns that have a substantial difference in prevalence between libraries and applications. The differences are not statistically significant ($p$-value $>0.05$). We can see that the variance of each distribution is high, which implies that the results may be due to the nature of each project instead of the project type, i.e., library or application.

**Generic and non-generic catch blocks.** Generic exceptions is an anti-pattern by itself, while some other anti-patterns, e.g., *Dummy Handler*, may be related to Generic catch blocks. We identified that there exists a significant difference (Wilcoxon Rank Sum test, $p$-value $<0.05$) between generic and non-generic catch regarding anti-patterns. Generic catch is a sign of developers' lack of knowledge on the possible exception(s), which explains the reason why developer may not know how to handle the exception but only log the exception instead (*Dummy Handler*). On the other hand, since generic catch may cover all possible exceptions from a try block, the chance of having *Unhandled exception* anti-pattern is smaller. Yet, such exception handling may mix critical issues with minor issues by only superficial handling strategies (like *Dummy Handler*), which may cause catastrophic failures of the software.

**Runtime and non-runtime exceptions.** Software is expected to recover more from non-runtime exceptions than runtime exceptions [13]. We compare anti-patterns detected with runtime and non-runtime exceptions in non-generic catch blocks, since generic exceptions are typically non-runtime. We find significant differences (Wilcoxon Rank Sum test, $p$-value $<0.05$) for *Destructive Wrapping, Incomplete Implementation and Throw within Finally*, only in Java projects. In all of those, the percentage of affected catch blocks is lower for runtime exceptions. Java does not force developers to handle runtime exceptions. Therefore, they are handled only if developers well understand the runtime exceptions, leading to fewer anti-patterns.

*F. Threats to validity*

**External validity.** Our findings may not generalize for other software, other programming languages or commercial software. **Internal validity.** Our study may not cover all possible anti-patterns. We selected anti-patterns based on the current research in the subject. Some anti-patterns that either 1) are out of the scope of exception handling (yet still mentioned in related work), 2) require heuristic to detect or 3) are not well explained in details in related work, are not included. Missing necessary documentation may also impact the identification of anti-patterns. **Construct validity.** The results in our study are based on catch blocks and throws statements. There may be other ways to measure the exception handling anti-patterns and their prevalence.

TABLE III
PERCENTAGE OF AFFECTED CATCH PER PROJECT PER ANTI-PATTERN.

| | Project | Flow | | | | Handler | | | | | | | | | | | | | | # Catch |
|---|---|---|---|---|---|---|---|---|---|---|---|---|---|---|---|---|---|---|---|---|
| | | Over-catch | Over-catch and Abort | Unhandled Exceptions | Unreachable Handler | Catch and Do Nothing | Catch and Return Null | Catch Generic | Destructive Wrapping | Dummy Handler | Ignoring Interrupted Exception | Incomplete Implementation | Log and Return Null | Log and Throw | Multi-Line Log | Nested Try | Relying on getCause() | Throw within Finally | | |
| C# | Glimpse | 33.33% | 0.00% | 12.28% | 63.16% | 7.02% | 7.02% | 75.44% | 1.75% | 21.05% | 0.00% | 0.00% | 0.00% | 0.00% | 0.00% | 0.00% | 3.51% | 0.00% | | 57 |
| | Google API | 40.00% | 0.00% | 43.33% | 60.00% | 10.00% | 0.00% | 56.67% | 20.00% | 10.00% | 0.00% | 0.00% | 0.00% | 0.00% | 0.00% | 0.00% | 6.67% | 0.00% | | 30 |
| | OpenRA | 23.08% | 0.70% | 14.69% | 58.74% | 23.08% | 19.58% | 76.22% | 1.40% | 12.59% | 0.00% | 0.00% | 14.69% | 0.00% | 3.50% | 0.00% | 0.00% | 0.00% | | 143 |
| | ShareX | 65.10% | 0.00% | 8.50% | 24.63% | 11.14% | 1.47% | 90.62% | 1.47% | 30.79% | 0.00% | 0.00% | 0.59% | 0.29% | 1.76% | 1.17% | 0.29% | 1.47% | | 341 |
| | Sharp D. | 32.17% | 0.09% | 40.38% | 45.75% | 18.30% | 10.00% | 45.75% | 4.15% | 13.40% | 0.00% | 0.38% | 3.30% | 0.09% | 11.79% | 8.96% | 0.66% | 0.66% | | 1,060 |
| | SignalR | 17.14% | 0.95% | 10.48% | 74.29% | 6.67% | 9.52% | 80.00% | 0.00% | 13.33% | 0.00% | 0.00% | 3.81% | 0.95% | 4.76% | 0.00% | 0.00% | 0.95% | | 105 |
| | Umbraco | 43.09% | 0.00% | 16.10% | 36.10% | 10.57% | 6.67% | 84.23% | 4.72% | 17.07% | 0.00% | 0.16% | 1.46% | 0.16% | 1.79% | 1.46% | 1.14% | 0.00% | | 615 |
| Java | Apache ANT | 31.26% | 0.09% | 69.80% | 6.94% | 11.76% | 3.34% | 17.56% | 37.14% | 5.09% | 2.81% | 0.26% | 0.53% | 0.26% | 1.67% | 5.79% | 0.35% | 14.05% | | 1,139 |
| | E. JDT Core | 11.72% | 0.24% | 69.06% | 11.78% | 31.24% | 11.18% | 3.14% | 4.71% | 7.25% | 1.09% | 0.06% | 0.48% | 0.06% | 0.06% | 2.36% | 0.18% | 8.22% | | 1,655 |
| | Elasticsearch | 24.26% | 0.00% | 24.51% | 24.51% | 10.54% | 4.17% | 33.82% | 31.62% | 8.09% | 3.43% | 0.00% | 0.98% | 0.00% | 1.23% | 4.41% | 0.98% | 3.19% | | 408 |
| | Guava | 19.87% | 0.00% | 27.44% | 37.22% | 4.73% | 10.09% | 26.50% | 24.61% | 5.05% | 7.89% | 0.32% | 0.95% | 0.00% | 0.32% | 0.95% | 6.94% | 10.73% | | 317 |
| | H. Common | 25.00% | 0.44% | 53.41% | 16.26% | 4.90% | 3.85% | 18.97% | 29.55% | 9.70% | 4.98% | 0.00% | 1.66% | 0.44% | 1.14% | 4.02% | 1.49% | 18.71% | | 1,144 |
| | H. HDFS | 12.46% | 0.17% | 41.30% | 30.55% | 3.24% | 1.37% | 2.22% | 34.13% | 5.29% | 11.43% | 0.00% | 1.02% | 0.68% | 1.88% | 1.19% | 0.85% | 4.44% | | 586 |
| | H. MapReduce | 15.80% | 0.00% | 49.32% | 16.08% | 3.00% | 7.08% | 13.35% | 41.69% | 8.17% | 14.99% | 0.00% | 3.54% | 0.54% | 1.09% | 3.27% | 0.82% | 30.25% | | 367 |
| | H. YARN | 15.57% | 0.39% | 43.75% | 20.01% | 2.55% | 6.80% | 12.69% | 30.80% | 10.01% | 4.91% | 0.13% | 1.70% | 0.26% | 1.64% | 2.62% | 1.05% | 35.71% | | 1,529 |
| | Spring | 28.55% | 0.00% | 48.46% | 25.51% | 7.95% | 3.00% | 29.99% | 40.03% | 7.82% | 0.74% | 0.00% | 1.56% | 0.04% | 0.91% | 2.17% | 1.78% | 4.82% | | 2,301 |

TABLE IV
PERCENTAGE OF AFFECTED THROWS PER PROJECT PER ANTI-PATTERN.

| | Apache ANT | E. JDT Core | Elastic search | Guava | Hadoop Common | Hadoop HDFS | Hadoop MapReduce | Hadoop YARN | Spring |
|---|---|---|---|---|---|---|---|---|---|
| Throws Kitchen Sink | 6.54% | 2.19% | 0.34% | 11.79% | 10.23% | 10.86% | | 3.77% | 9.62% | 8.21% |
| Throws Generic | 1.85% | 1.42% | 7.13% | 7.86% | 3.07% | 0.59% | 4.50% | 9.19% | 14.05% |
| # Throws | 1,622 | 1,686 | 1,782 | 509 | 4,495 | 1,538 | 1,221 | 4,146 | 5,856 |

TABLE V
DISTRIBUTION OF AFFECTED CATCH BLOCKS ACCORDING TO FLOW
ANTI-PATTERNS AND THE QUANTITY OF AFFECTED FLOWS.

| Affected Anti-patterns | Language | Quantity of Affected Flows | | | | | |
|---|---|---|---|---|---|---|---|
| | | 1 | 2 | 3 | 4 | 5 | >5 |
| *Unhandled Exceptions* | C# | 17% | 18% | 13% | 11% | 7% | 34% |
| and *Unreachable Handler* | Java | 33% | 16% | 11% | 8% | 5% | 28% |
| *Over-catch* and | C# | 16% | 16% | 12% | 10% | 6% | 31% |
| *Over-catch and Abort* | Java | 40% | 17% | 12% | 7% | 6% | 19% |

## IV. CONCLUSION

In this paper, we perform an empirical study using automatically detected 19 exception handling anti-patterns in 16 open source projects. We find that although all studied projects contain exception handling anti-patterns and every anti-pattern is detected in the source code, there exist only a small number of anti-patterns that are prevalent. These anti-patterns are often associated with multiple exception flows, making them more impactful and more difficult to address. With further investigation on the prevalence of anti-patterns, we find that the choice of programming languages may have a relationship to the introduction of anti-patterns. Our results suggest the need of in-depth study on exception handling practices. In particular, more user studies are required to further understand the choices of exception handling code and the introduction of exception handling anti-patterns. More importantly, future work should consider the impact of such exception handling code to assist in better resolution of exception handling anti-patterns and issues.